\documentclass[twocolumn,showpacs,pra]{revtex4}
\usepackage{amsmath}
\usepackage{amssymb}
\usepackage{amsfonts}
\usepackage{delarray}
\usepackage{graphicx}
\usepackage{float}
\usepackage{subfigure}
\newcommand{\bra}[1]{\langle #1|}
\newcommand{\ket}[1]{|#1\rangle}
\newcommand{\braket}[2]{\langle #1|#2\rangle}
\newcommand{\Tr}{\mathrm{Tr}}

\begin{document}

\title{Quantum Dynamics With Intrinsic Time Asymmetry and Indistinguishable Events}
\author{P. W.~Bryant}
\affiliation{Center for Complex Quantum Systems,
             Department of Physics,
             University of Texas at Austin, Austin, Texas 78712}

\date{\today}

\begin{abstract}
The extrinsic quantum mechanical arrow of time is understood to be
a consequence of the interaction between quantum systems and their environment.
A choice of boundary conditions for the Schr\"odinger equation results
in a different time asymmetry intrinsic to quantum mechanical dynamics
and independent of environmental interactions.
Correct application of the intrinsically asymmetric dynamics, however,
leads unavoidably
to predictions of the experimental signatures of the extrinsic arrow of time.
We are led to a new, model-independent mechanism for quantum
decoherence.
We need not invoke a master equation or a phase-destroying, non-Hermitian
Hamiltonian operator.
As an application, we calculate predictive probabilities
for the decoherence measured in Rabi oscillations experiments.
We can also show that a previously puzzling experimental result,
unexplained within the formalism of the quantum master equation,
is in fact expected and is the measurable consequence of the indistinguishability
of separate, uncontrolled interactions between systems and their environment.
\end{abstract}

\maketitle

\section{Introduction}
One cannot overstate
the practical importance of understanding the mechanisms responsible for
quantum decoherence.
The remarkable emergence of quantum engineering and the pursuit of quantum
computation have resulted in the creation of entire
industries relying upon the minimization of the effects of decoherence.

Less practical but perhaps more significant is the insight to be gained into
the theory of fundamental processes.
One experimental signature of the quantum mechanical arrow of time
is the decoherence measured for quantum mechanical systems.
A natural explanation for this observed arrow
for quantum systems, with their intrinsically reversible, unitary time evolution,
has been sought by many
investigators~\cite{wheeler_zurek_quantum_1983,schulman_times_1997,zeh_physical_2007}.
An arrow of time is usually accommodated within quantum theory
by adding to the unitary dynamics the effects of measurement by an
external system~\cite{von_neumann_meas_irrevers}.
This approach can be put in the form of a master equation,
which leads to predictions of decoherence and an irreversible time evolution
generated by a non-Hermitian,
dissipative operator~\cite{lindblad_generators_1976}.

There is, however, another type of time asymmetry for quantum mechanical
time evolution.
A choice of boundary conditions for the dynamical equations
leads to asymmetric time evolution described
by a semigroup and intrinsic to the dynamics of quantum state
vectors~\cite{physica_a_1997}.
This is the time asymmetry exhibited by the decay of metastable systems
and resonances, and can be thought of as existing
independently of interaction with the
environment~\cite{bohm_antoniou_kiel_arrow_of_time}.
A natural relationship between the extrinsic arrow and the intrinsic
time asymmetry has not previously been uncovered.

In classical physics, an analogy to this intrinsic time asymmetry
is the radiation arrow of time, which results from a choice of boundary
conditions~\cite{sommerfeld_partial_1949}.
An analogy to the extrinsic arrow of time is the thermodynamic arrow, which
is related to the statistics of configurations.
There once was even a famous argument over which was more fundamental,
resulting only in an agreement to disagree~\cite{einstein_ritz_disagree}.

After sketching briefly in Section \ref{sec:boundary_conditions}
how to find asymmetric time evolution
described by a semigroup, we will demonstrate in Section \ref{sec:consequences}
that the correct application of the resulting theoretical tools
requires one to distinguish between
time \textit{coordinates} and time evolution \textit{parameters}.
Such a distinction is not made when one works with the standard formalism
of the quantum master equation.
It is, however, the only requirement for a model-independent
mechanism predicting decoherence and the measurable signatures
of an extrinsic arrow of time.
We need not invoke a phase-destroying
master equation with a non-Hermitian Hamiltonian operator.
Nor do we require distinction between what is microscopic and what
is macroscopic.
The implication is that quantum mechanical time asymmetry is intrinsic to 
the dynamics, and that the extrinsic quantum mechanical arrow of time appears as a
consequence.

We do not suggest this new approach only for academic interest.
As an application, in Section \ref{sec:ap}
we successfully derive a predictive probability
for the measured decoherence of a 
quantum system undergoing Rabi oscillations.
We correctly match one result that has been particularly puzzling
because investigations with
the quantum master equation have failed to provide an explanation.
We can conclude that the previously puzzling, measured result is a consequence of
the indistinguishability of separate, uncontrolled interactions between quantum
systems and their environment.

Finally, in Section \ref{sec:qd}
we suggest that the theoretical basis of our approach can provide a
foundation for the natural incorporation of
dynamics into the theory of quantum mechanics.

\section{Boundary Conditions}
\label{sec:boundary_conditions}
Let us first illustrate how a choice of boundary conditions for the
Schr\"odinger equation,
\begin{equation}
\label{schrod_eqn}
i\hbar\frac{\partial}{\partial t}\phi=H\phi,
\end{equation}
can lead to time asymmetry \textit{intrinsic} to state vectors.
For simplicity, consider the time evolution of systems in
a pure state represented by a vector $\phi$
(or density operator $\rho=\ket{\phi}\bra{\phi}$.)
Let the observable be represented by a projection operator $\Lambda=\ket{\psi}\bra{\psi}$.
This section contains only a sketch, and for details the interested reader is 
referred to~\cite{bohm_annals_of_physics_small_preprint,
gadella_hardy_rhs,fortschritte_der_physik_1} and 
and the numerous references therein.

The Schr\"odinger equation is a differential equation for
the state vectors that represent physical systems.
When one is faced with a differential equation, the first step is to choose
boundary conditions.
Stone and von Neumann showed~\cite{stone_1932,von_neumann_1932} that, if
possible solutions, $\phi$, are chosen from the Hilbert space, $\mathcal{H}$,
then the solution of \eqref{schrod_eqn} is given by
\begin{equation}
\label{state_evolution}
\phi(t)=e^{-\frac{i H t}{\hbar}}\phi_0\,,\qquad-\infty<t<\infty,
\end{equation}
where $t$ parametrizes evolution in time.
Constraining the possible solutions by requiring $\phi\in\mathcal{H}$
is no different from constraining the solutions of any differential
equation by enforcing the proper boundary conditions.

The time evolution for states in \eqref{state_evolution} can
be described by the unitary group
\begin{equation}
\label{state_group}
U^\dagger(t)=e^{-\frac{i H t}{\hbar}}\,,\qquad -\infty<t<\infty.
\end{equation}
The group product is
\begin{equation}
U^\dagger(t_1)U^\dagger(t_2)=U^\dagger(t_1+t_2).
\end{equation}
For every evolution, $U^\dagger(t)$, there exists the inverse, $U^\dagger(t)^{-1}$,
given by
\begin{equation}
\label{eq:inverse}
U^{\dagger}(t)^{-1}=U^\dagger(-t).
\end{equation}
Because of the existence of \eqref{eq:inverse},
the time evolution of quantum state vectors is said to be intrinsically
symmetric in time.

The Schr\"odinger equation \eqref{schrod_eqn} is of course
unassailable, but the choice of
boundary conditions is apparently not.
One often chooses different boundary conditions
for \eqref{schrod_eqn}, though usually without realizing it.
When using Dirac kets ($\ket{E}$, $\ket{\vec{x}}$, etc.) with continuous spectra 
of eigenvalues,
one has chosen $\phi\in\mathcal{S}\subset \mathcal{H}$,
where the Schwartz space, $\mathcal{S}$, is a
subset of the Hilbert space~\cite{bohm_textbook}.

It has lately been realized that requiring further restrictions on the possible
solutions of \eqref{schrod_eqn} results in a time evolution for state
vectors subtly but fundamentally different from \eqref{state_evolution}.
For a variety of reasons stemming mainly from scattering theory and 
the theory of resonances and decaying states~\cite{bohm_bryant_quantal_2008},
a good choice of boundary conditions is 
\begin{equation}
\label{eq:hardy_axiom_1}
\textrm{set of possible states }\{\phi\} = \Phi_-\subset\mathcal{H}\subset \Phi^\times_-,
\end{equation}
where $\Phi_-$ denotes the Hardy space of the lower complex
semiplane, and $\Phi^\times_-$ is its dual~\endnote{
Specifically, the energy wave functions,
$\phi(E)$, are chosen to be smooth functions that can be analytically
continued into the lower complex energy plane.
Without going into too much detail, 
$\{\phi(E)\equiv \braket{E}{\phi}\}=(H^2_-\cap\mathcal{S})\vert_{\mathbb{R}_+}$,
where $H^2_-$ is the Hardy function space in the lower complex plane, 
and $\mathcal{S}$ is the Schwartz space~\cite{gadella_hardy_rhs}.}.
Note that, in practice, \eqref{eq:hardy_axiom_1} is not a limiting
restriction.
Any state vector in the Hilbert space can be approximated 
with arbitrary precision by state vectors in the Hardy
space~\cite{bohm_monograph_1989}.

With this new choice \eqref{eq:hardy_axiom_1} for the boundary conditions,
the solution of the Schr\"odinger equation
is~\cite{bohm_annals_of_physics_small_preprint}
\begin{equation}
\label{asym_states}
\phi(t)=e^{-\frac{i H^\times t}{\hbar}}\phi_0\,,\qquad 0\leq t<\infty.
\end{equation}
Note the lower bound on the time.
This is no longer time evolution given by the unitary group of \eqref{state_group}.
The time evolution is rather given by the \textit{semigroup} operator \endnote{
The operator notation $A^\times$ signifies that the operator is an extension of a Hilbert space
self-adjoint operator, $A=A^\dagger$, onto the space dual to the space of state vectors.
We will often drop it in what follows.},
\begin{equation}
\label{semigroup_states}
U^\times(t)=e^{-\frac{i H^\times t}{\hbar}}\,,\qquad 0\leq t<\infty.
\end{equation}
While the product of two elements is still defined by
\begin{equation}
U^\times(t_1)U^\times(t_2)=U^\times(t_1+t_2),
\end{equation}
being a semigroup means that the inverse, $U^{\times}(t)^{-1}$, of an element,
$U^\times(t)$ with $t>0$, does not exist.
In other words, one can choose only positive values for the time evolution parameter,
and one cannot drive time evolution to earlier times.
Evolution given by the semigroup is thus intrinsically asymmetric in time.

This time asymmetry has been decreed by our choice of boundary conditions.
We have said nothing deep about the nature of time;
we have only made a phenomenologically sound choice.
Furthermore, this time asymmetry is fundamentally different
from any asymmetry consequent only to the interaction of a system
with its environment.

\section{Consequences}
\label{sec:consequences}
The intrinsic time asymmetry has a handful of very profound consequences.
Perhaps most remarkable is that, when one correctly uses
state vectors with the semigroup time evolution~\eqref{semigroup_states}
to represent physical systems,
the signature of the extrinsic arrow of time arises \textit{unavoidably}
and in a manifestly model-independent fashion.
To understand why, let us list some of the consequences
of our choice of boundary conditions.

\subsection{Preparation time}
As one must always do, we shall distinguish between theoretical objects
and the physical systems they represent.
In the theory, there is now a special time, $t=0$,
belonging to the state vectors that represent physical systems.
It should come as no surprise that,
by appealing to the Born probability, we can identify this special
time of the theory uniquely as the time corresponding to the preparation
of physical systems.
In the Schr\"odinger picture,
the Born probability to find the observable $\Lambda=\ket{\psi}\bra{\psi}$ in 
the state $\rho(t)=\ket{\phi(t)}\bra{\phi(t)}$ is
\begin{equation}
\label{asymmetric_born}
\mathcal{P}_{\Lambda}\big(\rho(t)\big)=\Tr\big(\Lambda\,\rho(t)\big)
=|\braket{\psi}{\phi(t)}|^2,
\qquad 0\leq t<\infty.
\end{equation}
Here the time evolution comes from \eqref{asym_states},
and the calculated probability compared to measurements
is thus defined only for $0\leq t<\infty$.

In an experiment, for any physical system
represented by a state vector,
there is also a special time: the preparation time, $t_{prep}$.
It is the time at which the physical system
has been prepared such that it is representable by, say, $\phi(t)$.
It is also the time after which a detector can possibly register an observable:
\begin{eqnarray}
\label{eq:state_prep}
&\textrm{the system represented by }\phi(t) \nonumber \\
&\textrm{is prepared at }t=t_{prep},
\end{eqnarray}
and
\begin{eqnarray}
\label{eq:obs_prep_meas}
&\textrm{the observable represented by }\ket{\psi}\bra{\psi} \nonumber \\
&\textrm{is registered
at }t\geq t_{prep}.
\end{eqnarray}

Comparing the Born probability, \eqref{asymmetric_born},
with the phenomenological statements,
\eqref{eq:state_prep} and \eqref{eq:obs_prep_meas},
we identify the semigroup time, $t=0$, of
\eqref{asymmetric_born},
and thus \eqref{asym_states}, with the
preparation time, $t_{prep}$,
of the physical systems represented by state vectors.
Though this is very basic, it is also unusual.
But it follows necessarily from our choice 
of boundary conditions \eqref{eq:hardy_axiom_1}.

\subsection{Ensembles}
\label{sec:ensembles}
Quantum mechanics is a probabilistic theory, and the results of quantum
mechanical calculations are Born probabilities and
are to be compared to ensemble averages
over experimental results~\cite{dirac_qm_book}.
Quantum mechanical state vectors are the theoretical objects
representing physical systems, and so, in practice,
those state vectors represent ensembles of physical systems.

Above, we have identified the semigroup time from the theory,
$t=0$, with the preparation time, $t_{prep}$, of physical systems.
Therefore, if $\phi(t)$ is to represent all of the identically prepared
physical systems present in an experiment, as is always the assumption, then
we are forced to establish the following rule:
\begin{description}
\item[Rule:] Every physical system,
at the moment it is prepared and thus representable
by the state vector $\phi(t)$, is represented by that vector at the time
zero, $\phi(t=0)$,
\textit{regardless of the (coordinate) time in the lab at which the preparation occurs}.
\end{description}

This rule requires some commentary.
First, we are not concerned here with the details of how the state of a physical
system changes from something represented by, for example, $\phi_{before}(t)$,
into a state represented by $\phi_{after}(t)$.
If the reader is bothered by the ``moment'' in time referred to above, then this
rule should be considered effective or pragmatic.
To the best of the author's knowledge, however, up to experimental precision,
when dealing with discrete states
quantum jumps have been observed, and smooth transitions from state to state
have not~\cite{nagourney_dehmelt_shelved_1986,bergquist_qjumps_1986,
sauter_toschek_qjumps_1986,peik_qjumps_1994}.

Second, though this rule is not always followed in the theory, especially
when there is no distinguishable time available \eqref{state_evolution},
it is followed in the analysis of experimental data.
For example, consider the remarkable quantum jumps experiments
in~\cite{nagourney_dehmelt_shelved_1986,bergquist_qjumps_1986,
sauter_toschek_qjumps_1986,peik_qjumps_1994}.
Physicists repeatedly measure lifetimes of a metastable state of a single atom
as durations in laboratory time.
By themselves,
single measurements are useless, so the resulting
lifetime of the metastable state is found
from the ensemble average of as many measured durations as possible
(or by fitting them to a decaying exponential.)
The durations in time start at $t=0$, regardless of when, according to the clock on the
laboratory wall, any single atom was prepared to be in its metastable state.

Based on an interpretation of the master equation formalism,
it has been argued~\cite{zeh_physical_2007} that in these quantum jumps experiments the 
jumps and the ensemble are illusory, and that the data result rather from measurements
of rapidly decohering parts of entangled wave functions that are continuously defined
in time.
Though we acknowledge different interpretations,
here we are only interested in how our theoretical objects represent physical systems.
With our approach, we must work with ensembles.

Note also that
the people performing these experiments on
single atoms, and analyzing the data, do in fact treat their data as ensembles.
The same is true for those performing the experiments on single
systems undergoing Rabi
oscillations~\cite{meekhof_rabi_1996,brune_rabi_1996,petta_coherent_2005},
that we will study later.
In~\cite{meekhof_rabi_1996}, single physical systems are prepared and then undergo
oscillations for a controlled length of time before an active measurement occurs.
As always, any length of time begins at $t=0$, and the measured result is an ensemble average
over approximately $4000$ repetitions for every length.
Indeed, the use of ensembles is not limited to one type of experiment.
In scattering experiments, for instance, results are presented as histograms,
which are pictures of ensembles.

Working with ensembles is therefore justified phenomenologically.
In the end, however, these arguments for justification are only academic.
In Section~\ref{sec:compare}
we will show that, by following the rule above, we can calculate
predictive probabilities matching experimental results.
One of the measurements we match has been particularly puzzling,
because, for it, investigations using the master equation formalism have not succeeded.

\subsection{Coordinates and Parameters}
Because of our choice of boundary conditions, we must distinguish
between time coordinates and time parameters.
Let us label time coordinates with a tilde: $\tilde{t}$.
Time coordinates are absolute times indicated by laboratory clocks.
Experimental measurements are performed in coordinate time,
but those results are independent of any \emph{specific value} of the
time coordinate.
This is a statement of macroscopic time translation invariance.
Most importantly, time coordinates do not parametrize the time
evolution of the state vectors representing physical systems.

Let us continue to label time evolution parameters with the letter $t$.
Time evolution parameters are found in the dynamical equation \eqref{schrod_eqn}
and its solutions,
\eqref{asym_states} and \eqref{semigroup_states}.
They do parametrize the time evolution of quantum mechanical state vectors.
They always correspond to durations, and,
in general, there is no time
translational invariance for time parameters.

Distinguishing coordinates from parameters is not new.
In relativistic quantum mechanics, they are
often represented by different symbols.
But in non-relativistic quantum mechanics with time symmetric boundary conditions, 
one does not usually distinguish between the two.
Perhaps because the domains of definition of $t$ and $\tilde{t}$ overlap,
one feels free to substitute $\phi(t)$ with $\phi(\tilde{t})$---or
$\rho(t)$ with $\rho(\tilde{t})$ if one works with density 
operators---thus implicitly making the identification
\begin{equation}
\label{eq:subin}
t\Leftrightarrow \tilde{t}.
\end{equation}
It is precisely this identification that we can no longer make when using
time asymmetric boundary conditions.

It is obvious that \eqref{eq:subin} is problematic when we consider that 
possible values of
$t$ and $\tilde{t}$ are no longer even chosen from the same set:
\begin{equation}
\label{eq:domains}
t\in [0,\infty)\qquad\textrm{and}\qquad\tilde{t}\in (-\infty,\infty).
\end{equation}

Consider also the rule established in the previous section.
For a state vector, $\phi(t)$, representing an ensemble of systems,
the theoretical preparation time, $t=0$, corresponds in general
to an ensemble of time coordinate values, $\{\,\tilde{t}_{prep}\,\}$,
that mark the times on laboratory clocks at which members of the ensemble
were prepared.
(In some experiments, this ensemble of times is recorded and
referred to as the ``time-stamps.'')
When the theoretical state vectors have time evolution that is intrinsically
asymmetric, there can be no one-to-one correspondence
between the parametric value $t=0$
and values of the laboratory time coordinate, $\tilde{t}$.
Making the identification in \eqref{eq:subin} is 
thus no longer possible.

\subsection{Measurements}
Experimental measurements are, of course, performed in the
coordinate time of the laboratories.
Therefore, dynamical measurements result in time series that are functions
of the coordinate time, $\tilde{t}$.
The results of dynamical calculations,
which are solutions of the Schr\"odinger equation
and are given by \eqref{asym_states},
are instead functions of the time evolution parameter, $t$.
To compare theory with experiment, then, we must
determine the theoretical Born probabilities also as functions
of the coordinate time, $\tilde{t}$:
\begin{eqnarray}
\label{eq_bornshift}
\mathcal{P}_{\Lambda}\big(\rho(t)\big)=\Tr\big(\Lambda\,\rho(t)\big)&\rightarrow&
\mathcal{P}_{\Lambda}\big(\rho(\tilde{t})\big)=\Tr\big(\Lambda\,\rho(\tilde{t})\big)\,\,\, \\
\textrm{Calculated}\qquad & &\qquad \textrm{Measured} \nonumber
\end{eqnarray}
Equivalently, we can make the transformation
\begin{equation}
\label{eq_rhoshift}
\rho(t)\rightarrow \rho(\tilde{t}).
\end{equation}

Here $\rho(t)$ represents the state of the quantum
systems as a function of the time evolution parameter, $t$.
On the other hand, $\rho(\tilde{t})$ represents the state of
the quantum systems as a function of the coordinate time of the
laboratory where measurements are performed.
Measured results are thus predicted by $\mathcal{P}_{\Lambda}\big(\rho(\tilde{t})\big)$
rather than by $\mathcal{P}_{\Lambda}\big(\rho(t)\big)$.
We shall hereafter refer to $\mathcal{P}_{\Lambda}\big(\rho(\tilde{t})\big)$
as the ``predictive probability.''

The statements in \eqref{eq_bornshift} and \eqref{eq_rhoshift}
are the most important results of this paper.
Clearly they are independent of any model,
and they represent a change from the standard theoretical
approach~\cite{petruccione_open_quantum_systems},
which will be briefly described later.
The transformations in \eqref{eq_bornshift} and \eqref{eq_rhoshift}
are required if one chooses intrinsically time asymmetric boundary conditions
and wishes to compare theoretical calculations with the results of experiments.
As we shall see next,
predictions of the experimental signatures of the extrinsic arrow
of time necessarily follow.
We have thus linked
the intrinsic time asymmetry to the measured, extrinsic 
quantum mechanical arrow of time.

\subsection{General Hypotheses}
In the next section we will deduce a simple, specific 
transformation scenario for \eqref{eq_bornshift} and use
it to calculate a predictive probability,
$\mathcal{P}_{\Lambda}\big(\rho(\tilde{t})\big)$.
Let us first develop general hypotheses.

Imagine an ensemble of physical systems prepared by a physicist
to undergo a dynamical process followed at the end by an active measurement.
(And, as we now know, not all systems must be present in the lab
simultaneously.)
Because an ensemble cannot be perfectly isolated from its environment,
at times after preparation and before active measurement, a number of systems
from the ensemble
suffer environmental perturbations in which they are passively measured.
Passive measurement is the physical process often referred to in theory
as ``collapse of the wave function'' to the eigenstates of observables.
And because the measurements
are passive, physicists gain no information regarding which
eigenstates are chosen by nature.
Of course, measuring a system to be in a state
is equivalent to preparing that system to be in that state.
Passive measurements are therefore equivalent to passive preparations.

\begin{figure}[htp]
  \begin{center}
    \subfigure[ Before interference event.  
    $\rho(\tilde{t}-\tilde{t}_{prep})\Leftrightarrow 
    \rho(t)$.]{\label{before_samp}\includegraphics[scale=0.5]{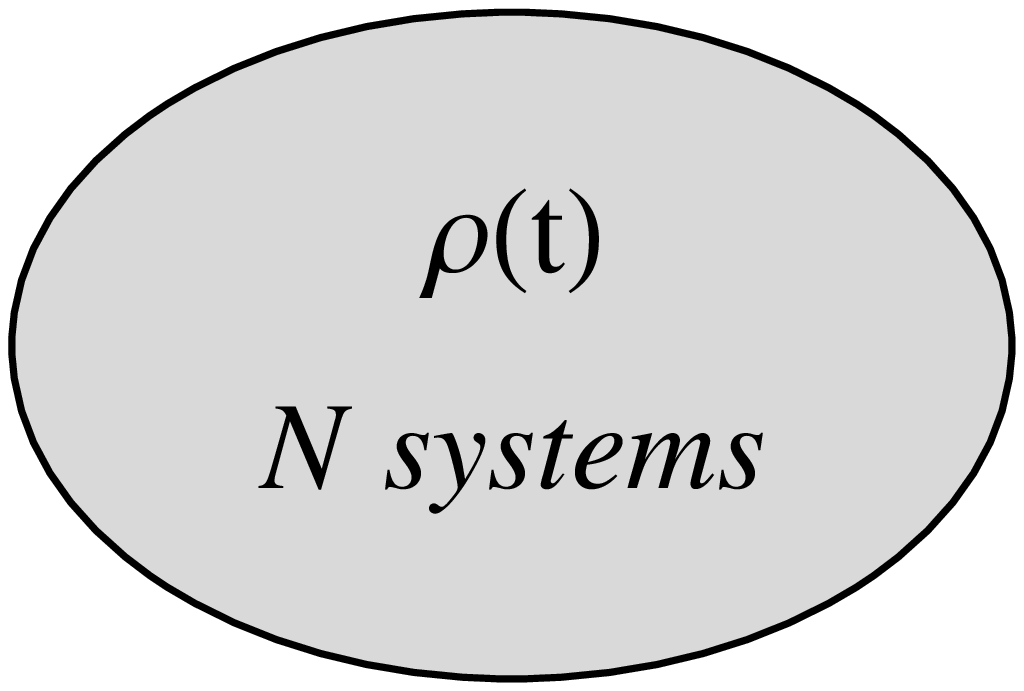}}\hfill
    \subfigure[ After interference event.  
    $\rho(\tilde{t}-\tilde{t}_{prep})\Leftrightarrow 
    F\big(\rho(t),\rho^\prime(t^\prime),N,m,\tilde{t}_1-\tilde{t}_{prep}\big)\neq \rho(t)$.
    The preparation times $t=0$ and $t^\prime =0$ \emph{do not} correspond
    to the same value of the time
    coordinate, $\tilde{t}$.]{\label{after_samp}\includegraphics[scale=0.58]{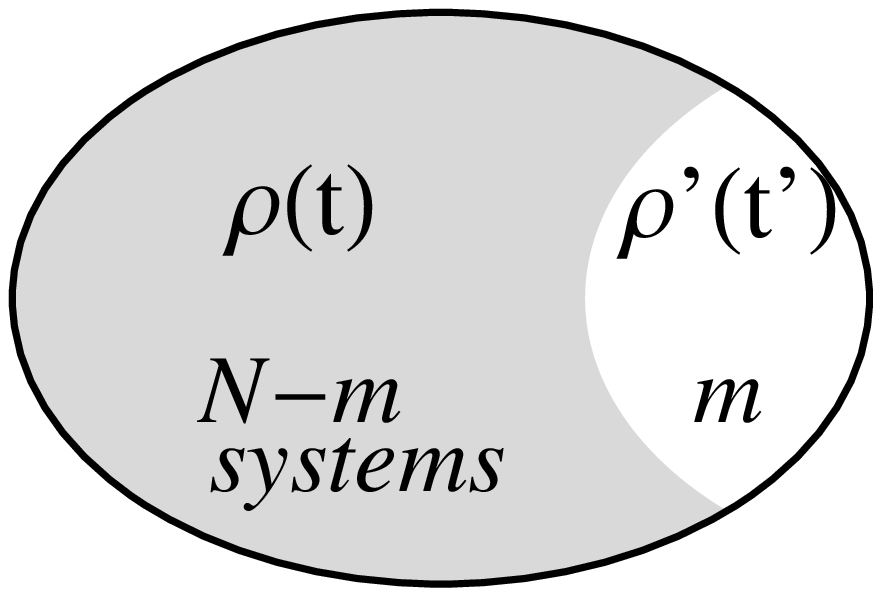}}
  \end{center}
  \caption{Ensemble of \emph{distinguishable} systems before and after a single interference event.
  Such a diagram for indistinguishable systems would be difficult to draw.
  (See Section~\ref{sec:prob_indist}.)}
  \label{fig:ensemble}
\end{figure}

We will break this scenario into three steps.  A schematic
can be seen in Figure~\ref{fig:ensemble}.
Though we will eventually work with indistinguishable systems,
for simplicity, we refer here to distinguishable systems.
\begin{description}
\item[Before perturbation:] Let us assume that $N$ systems are prepared
initially to be in the pure state represented by $\rho(t)$.
While these systems evolve in time prior to any environmental interference,
they continue to be described by $\rho(t)$, and the transformation
in \eqref{eq_rhoshift} is trivial:
$\rho(\tilde{t}-\tilde{t}_{prep}) = \rho(t)$.
Here $\tilde{t}_{prep}$ refers to the value of the time coordinate
corresponding to active preparation.
It is the time on the clock, at which the experiment begins.
This is pictured in Figure~\ref{before_samp}.
\item[Perturbation occurs:] At some time, $\tilde{t}_1$,
prior to active measurement,
some number $m$ of the systems suffer environmental perturbation
such that they are passively prepared to be in a new, generally
mixed state.
The unperturbed systems, numbering $N-m$, are unaffected.
\item[After perturbation:] The $N-m$ unperturbed systems remain
in the pure state represented by $\rho(t)$.
The $m$ systems that suffered perturbation have been passively
prepared to be in some mixed state represented by $\rho^\prime(t^\prime)$.
The time evolution of $\rho^\prime(t^\prime)$ is
determined, as always, by the Schr\"odinger equation,
the initial conditions,
and the appropriate, self-adjoint Hamiltonian operator.
Because a subset of systems is no longer represented by
$\rho(t)$, the transformation in \eqref{eq_rhoshift} is not trivial.
This is pictured in Figure~\ref{after_samp}.
\end{description}

After the first occurrence of environmental interference, the density
operator as a function of time in the laboratory,
$\rho(\tilde{t}-\tilde{t}_{prep})$, is in general some operator valued function
$F\big(\rho(t),\rho^\prime(t^\prime),N,m,\tilde{t}_1-\tilde{t}_{prep}\big)$.
Furthermore, the time parameters $t$ and $t^\prime$ have no
trivial correspondence.
Because of the rule established in Section~\ref{sec:ensembles},
the parameter value $t=0$ corresponds to the time coordinate value of the
active preparation, $\tilde{t}_{prep}$, while the parameter
value $t^\prime = 0$ corresponds to the time coordinate value
of the passive preparation, $\tilde{t}_{prep}+\tilde{t}_1$:
\begin{eqnarray*}
t=0 & \Leftrightarrow & \tilde{t}_{prep} \\
t^\prime=0 & \Leftrightarrow & \tilde{t}_{prep} + \tilde{t}_1
\end{eqnarray*}

This hypothetical scenario is quite simple.
We must always take care, however, not to break the rules of nature.
For example, in the next section we will face the interesting complication
of dealing with an ensemble of indistinguishable systems suffering
indistinguishable interference events.

While the statements \eqref{eq_bornshift} and \eqref{eq_rhoshift} 
provide a model-independent mechanism for decoherence,
for a detailed study one needs a model for the
interaction of the environment with the physical systems of interest.
This is where model-independence ends. 
If experiments are well understood, however,
then there ought to be no adjustable parameters
in a given transformation scenario.
This last statement can be reversed.
An understanding of this mechanism for decoherence,
coupled with experimental measurements,
ought to provide a valuable diagnostic tool
for understanding how systems interact with an uncontrolled environment.

\section{Application:  Rabi Oscillations Experiments}
\label{sec:ap}
As a test of our approach, we shall study the non-dissipative
decoherence measured in systems undergoing Rabi oscillations.
Though to be understood fully some experiments will require detailed
transformation scenarios, for these experiments
a few simple assumptions will suffice
to predict the available experimental observations.

The theory of Rabi oscillations is well established~\cite{dodd_atoms_1991}.
Consider a two level system, with levels described by
the projection operators
$\ket{g}\bra{g}$ representing the ground state and
$\ket{e}\bra{e}$ representing the excited state.
When the system is prepared at $t=0$ to be in the excited state,
$\ket{e}$, and it is coupled to a correctly tuned
radiation field,
the Born probability at $t$ to find the system in the ground state,
$\ket{g}$, is
\begin{equation}
\label{born_rabi}
\mathcal{P}_{\ket{g}\bra{g}}\big(\rho(t)\big)= \textrm{sin}^2(\Omega t)=
\frac{1}{2}\big(1-\textrm{cos}(2 \Omega t)\big),
\end{equation}
where $\Omega$ is called the Rabi frequency.
(The probability to find a system in that state in which it was prepared
is of the form $\textrm{cos}^2(\Omega t)$.)
These expressions are calculated using the rules and equations
of quantum mechanics.
The $t$ in \eqref{born_rabi} is therefore
a time parameter rather than a time coordinate.

This is the story theoretically.
Experimental measurements, however, reveal decoherence because
real, physical systems interact with their environment.
As energy is lost from dissipative systems, the signature
of decoherence is a damping of the oscillations until, in the steady state,
all systems are found to be in the ground level, $\ket{g}$.

In this paper, however, we wish to consider three varied
but very clean and well-controlled
experiments~\cite{meekhof_rabi_1996,brune_rabi_1996,petta_coherent_2005}
in which the observed decoherence was found to be non-dissipative in
nature.
In these experiments, at large times the steady state probability
to find the system in either the excited or the ground state is $\frac{1}{2}$.
As explained in~\cite{bonifacio_intrinsic_deco_rabi_2001}, the 
steady state values of $\frac{1}{2}$ cannot be explained by invoking
dissipative mechanisms.

In one experiment~\cite{meekhof_rabi_1996},
internal levels of a $^9$Be$^+$ ion couple to
a harmonic binding potential.
Rabi oscillations occur between two of the coupled internal and vibrational
levels, and the oscillations between different sets of levels are measured.
In another experiment~\cite{brune_rabi_1996},
Rabi oscillations are observed between the
circular states of a Rydberg atom coupled to a field stored in
a high $Q$ cavity.
In the third experiment~\cite{petta_coherent_2005},
the Rabi oscillations are between the spin states of
two electrons in a double quantum dot.
This physical system may be useful for quantum computation and is
therefore of great interest.

In all of these experiments,
single systems are prepared and then undergo the dynamics individually.
Experimental results follow from ensemble averages over
the measurements made on these systems.
One makes the assumption, of course, that performing
multiple measurements on individual systems is equivalent
to performing one measurement on multiple, simultaneously
present but non-interacting systems.

In none of the experiments does the result agree with the
probability calculated in \eqref{born_rabi}.
Rather, the measured probability is fit by an appropriately damped sinusoid,
\begin{equation}
\label{damped_rabi}
\mathcal{P}_{\ket{g}\bra{g}}\big(\rho(\tilde{t})\big)=
\frac{1}{2}\big(1-e^{-\gamma \tilde{t}}\,\textrm{cos}(2 \Omega \tilde{t})\big),
\end{equation}
where $\gamma$ is an experimentally determined decay factor.
(In~\cite{petta_coherent_2005}, amplitude, offset, and phase are fit as well,
because the oscillations are measured also as a function of a swept detuning voltage.)
The decaying exponential in \eqref{damped_rabi} has been inserted to 
parametrize the rate of decoherence.
In two of the experiments~\cite{meekhof_rabi_1996,petta_coherent_2005},
the decay factor, $\gamma$, has been investigated and found
to depend on the Rabi frequency, $\Omega$.
In~\cite{meekhof_rabi_1996}, this dependence has been measured in detail.
It will be discussed later.

Our approach provides a very natural explanation for the
discrepancy between measurement and theory:
\emph{Dynamical measurements are made in the lab.}
The $\tilde{t}$ in \eqref{damped_rabi} is therefore a time coordinate.
The calculated Born probability in \eqref{born_rabi} is a function of the
time parameter, $t$.
It corresponds to the left hand side of \eqref{eq_bornshift}.
The measured probability in \eqref{damped_rabi} corresponds to the right hand side
of \eqref{eq_bornshift}.
To reconcile measurement with the theory, we must find the correct relationship between
the calculated
$\mathcal{P}_{\ket{g}\bra{g}}\big(\rho(t)\big)$
and the predictive
$\mathcal{P}_{\ket{g}\bra{g}}\big(\rho(\tilde{t})\big)$.

\subsection{Predictive Probability}
\label{sec:dist_dev}
To transform the calculated probability \eqref{born_rabi}
into the predictive functions \eqref{damped_rabi}
of the time coordinate, $\tilde{t}$, we must
assume a specific transformation scenario based on the interactions of systems
with their environment.
For now we will again treat the systems as distinguishable.
The simplest reasonable algorithm the author can conceive is:
\begin{enumerate}
\item Systems are actively prepared in the lab at $\tilde{t}=\tilde{t}_{prep}$.
Because the actual value of the time coordinate corresponding
to $\tilde{t}_{prep}$ is physically irrelevant, and
for simplicity of notation, we let $\tilde{t}_{prep}=0$ and write
\begin{equation}
\mathcal{P}_{\ket{g}\bra{g}}\big(\rho(\tilde{t})\big)
\equiv
\mathcal{P}_{\ket{g}\bra{g}}\big(\rho(\tilde{t}-\tilde{t}_{prep})\big).
\end{equation}
\item The physical systems in the ensemble can possibly
suffer environmental interactions at the times
$n\Delta\tilde{t}$, where $n=1,2,3\ldots$
As a result, some systems are passively prepared to be in either the ground
state or the excited state.
We shall call these interactions ``interference events.''
\item At every $n\Delta\tilde{t}$, there is some probability, $\lambda$,
for a member system to suffer perturbation and thereby to be prepared passively.
\end{enumerate}
In the second step, we have introduced the time scale, $\Delta\tilde{t}$,
of interaction with the environment.
In the third step we have introduced a parameter, $\lambda$,
to represent the susceptibility
of our physical systems to environmental interference.
We prefer, however, to work with the parameter
\begin{equation}
\eta\equiv(1-\lambda), \qquad 0\leq\eta\leq 1.
\end{equation}
The parameter $\eta$ is therefore the probability for a
system from the ensemble \emph{not to suffer} interference at one of the
times $n\Delta \tilde{t}$.
For a perfectly isolated system, $\eta=1$.

This scenario somewhat resembles the method of quantum Monte Carlo
trajectories~\cite{haroche_exploring_quantum_book},
in which quantum state vectors are allowed to evolve in small
time steps separated by quantum jumps to new state vectors
having new normalizations.
Though to maintain transparency we will choose to solve recursively
for a predictive probability, a Monte Carlo method indeed works
just as well for our scenario.
One important difference, however, is that,
in the method of trajectories, even \emph{between} the quantum
jumps, the state vectors evolve in time according to an effective,
non-Hermitian Hamiltonian,
\begin{equation*}
H_{eff}=H-i\hbar J, \qquad (\textrm{Master equation formalism})
\end{equation*}
where the operator $J$ depends on the character
of environmental interactions.
With such an evolution in time,
it can be shown~\cite{haroche_exploring_quantum_book}
that after very many simulated trajectories have been
averaged, the method of quantum Monte Carlo trajectories is
equivalent to solving the Lindblad form of the master equation.
Because of our choice of boundary conditions \eqref{eq:hardy_axiom_1},
we need not use effective, non-Hermitian
Hamiltonian operators in our scenario;
our method is not equivalent to solving the Lindblad form of the master equation.
As will be demonstrated in Section~\ref{sec:prob_indist},
we will also enjoy more freedom when we study dynamics.

Before we derive an equation, note that
our scenario is based on the repeated,
passive measuring of physical systems in an experimental ensemble.
Rather than involving probability amplitudes,
measurement involves probabilities directly.
We shall therefore concern ourselves with
calculating the predictive probability,
$\mathcal{P}_{\ket{g}\bra{g}}\big(\rho(\tilde{t})\big)$,
rather than deducing the density matrix as a function of the
time coordinate, $\rho(\tilde{t})$.

According to our simple transformation scenario,
in which subsets of an ensemble of physical systems
suffer environmental interference at the times $n\Delta\tilde{t}$,
we can write a very general formula
for the predictive probability,
$\mathcal{P}_{\ket{g}\bra{g}}\big(\rho(\tilde{t})\big)$:
\begin{equation}
\label{prob_written}
\mathcal{P}_{\ket{g}\bra{g}}\big(\rho(\tilde{t})\big) =
\begin{cases}
p_0(\tilde{t}) & 0\leq\tilde{t}< 1\,\Delta\tilde{t} \\
p_1(\tilde{t}) & 1\,\Delta\tilde{t}\leq\tilde{t}< 2\,\Delta\tilde{t} \\
\quad\vdots & \qquad\vdots \\
p_n(\tilde{t}) & n\Delta\tilde{t}\leq\tilde{t}< (n+1)\Delta\tilde{t} 
\end{cases}
\end{equation}

We will assume that all systems are initially prepared to be in the
excited state, $\ket{e}$.
Because no systems will have suffered environmental interference before
$\tilde{t}=1\,\Delta\tilde{t}$,
we have for the initial value
$p_0(\tilde{t})=\textrm{sin}^2(\Omega \tilde{t})$.
Then, at $\tilde{t}=1\,\Delta\tilde{t}$, the fraction $(1-\eta)$
of the ensemble
will be passively prepared to be in a new state, with new
initial conditions.
For the probability
after the first perturbation and before the second perturbation,
$p_1(\tilde{t})$, we can write
\begin{widetext}
\begin{equation}
\label{eq:first_rec}
p_1(\tilde{t})=\eta\,p_0(\tilde{t})+
(1-\eta)
\Big(\textrm{cos}^2\big(\Omega\,(\tilde{t}-1\Delta\tilde{t})\big)\,p_0(1\Delta\tilde{t})
+ \textrm{sin}^2\big(\Omega\,(\tilde{t}-1\Delta\tilde{t}) \big)\,
\big(1-p_0(1\Delta\tilde{t})\big) \Big).
\end{equation}
\end{widetext}

The equation \eqref{eq:first_rec} has a straightforward explanation.
The term $\eta\,p_0(\tilde{t})=\eta\,\textrm{sin}^2(\Omega\tilde{t})$ corresponds to
the unperturbed subset of the ensemble that passes the time $1\Delta\tilde{t}$
without suffering interaction with the environment.
The term multiplied by $(1-\eta)$ corresponds to the perturbed subset
of physical systems.
When the perturbation occurs at $\tilde{t}=1\Delta\tilde{t}$,
the probability for the perturbed systems to be found and therefore
prepared to be in the ground state is
\begin{equation*}
\label{pert_11}
\mathcal{P}_{\ket{g}\bra{g}}\big(\rho(1\,\Delta \tilde{t})\big)
=\textrm{sin}^2(1\,\Omega \Delta \tilde{t}).
\end{equation*}
The probability at $\tilde{t}=1\Delta\tilde{t}$
for the perturbed systems to be found in the excited state is
\begin{equation*}
\label{pert_12}
\mathcal{P}_{\ket{e}\bra{e}}\big(\rho(1\,\Delta \tilde{t})\big)
=1-\mathcal{P}_{\ket{g}\bra{g}}\big(\rho(1\,\Delta \tilde{t})\big)
=\textrm{cos}^2(1\,\Omega \Delta \tilde{t}).
\end{equation*}

According to the standard calculations for Rabi oscillations, if 
systems are prepared at $1\Delta\tilde{t}$, even passively,
to be in the ground state,
the probability to find them at a later time $\tilde{t}$ also in the ground
state is $\textrm{cos}^2\big(\Omega(\tilde{t}-1\Delta\tilde{t})\big)$.
With this shift in time, we explicitly follow the rule
from Section~\ref{sec:ensembles}, which states that the time parameter
describing the evolution of a state vector (and therefore the
Born probability) must be set to $t=0$ if it is to represent a physical
system at its moment of preparation~\endnote{Here we use the fact that
durations in parametric time are equivalent to durations in
coordinate time.}.
In \eqref{eq:first_rec} the term
\begin{equation*}
\textrm{cos}^2\big(\Omega\,(\tilde{t}-1\Delta\tilde{t})\big)\, p_0(1\Delta\tilde{t})
\end{equation*}
is thus the probability that systems are passively prepared
at $1\Delta\tilde{t}$ to be in the ground state, $\ket{g}$, and will also
be found again in the ground state at $\tilde{t}$, with
$1\Delta\tilde{t}\leq\tilde{t}<2\Delta\tilde{t}$.
Similarly, the third term in \eqref{eq:first_rec}
is the probability that systems are passively prepared
at $1\Delta\tilde{t}$ to be in the excited state, $\ket{e}$, and will
be found in the ground state at $\tilde{t}$.

\begin{figure}[htp]
  \begin{center}
    \subfigure[With $\eta=0.99$, we fit $\gamma/\Omega=0.05$.]
    {\label{rec1}\includegraphics[width=0.45\textwidth]{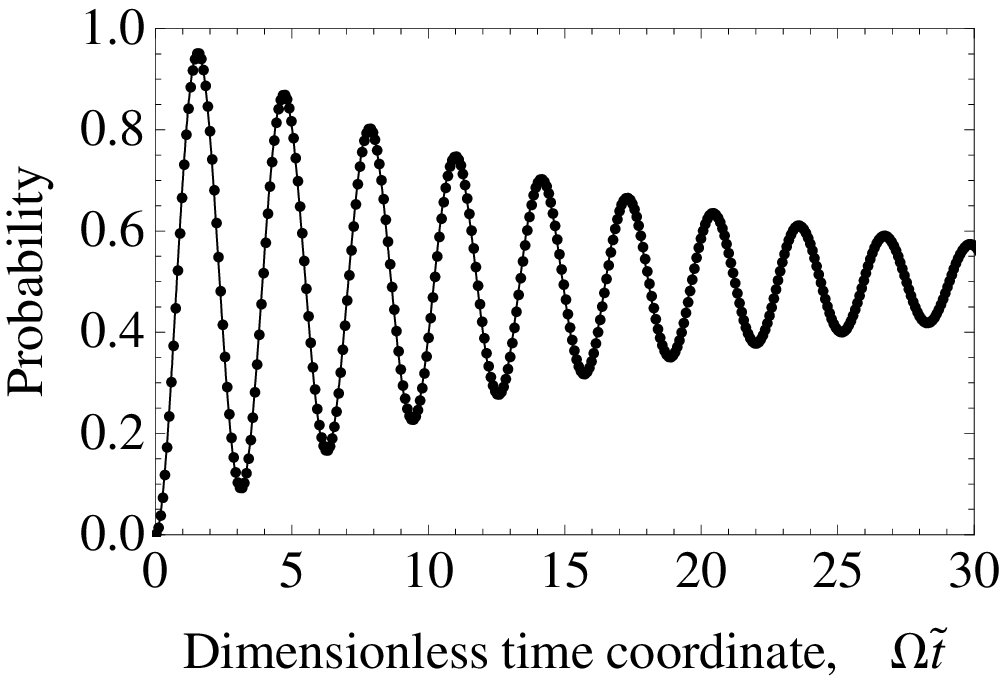}}\hfill
    \subfigure[With $\eta=0.997$, we fit $\gamma/\Omega=0.015$.]
    {\label{rec2}\includegraphics[width=0.45\textwidth]{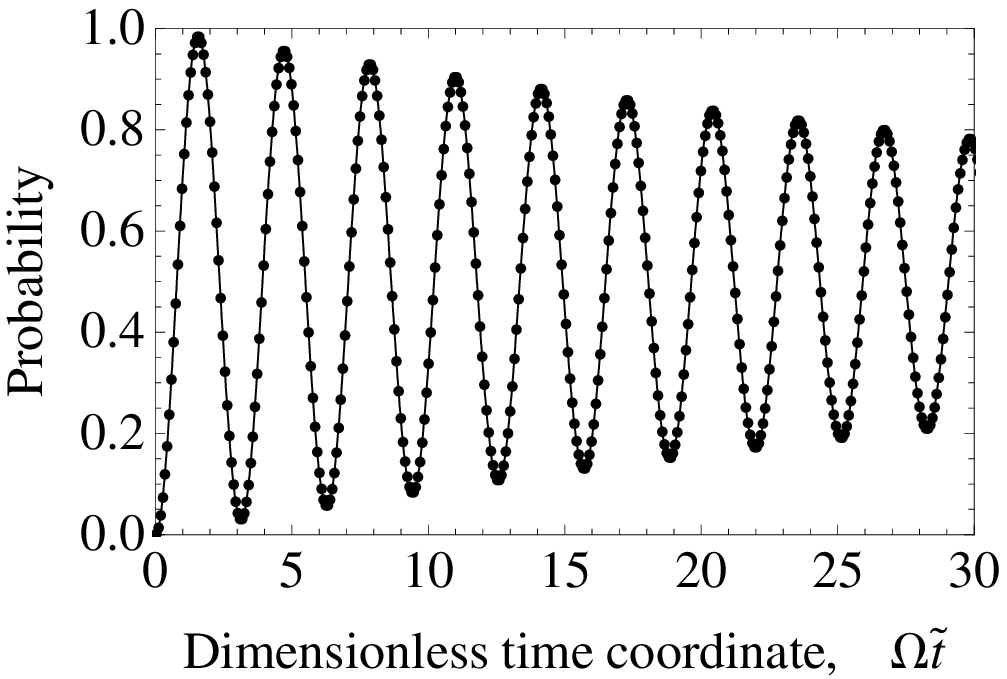}}\\
  \end{center}
  \caption{The predictive probability,
           $\mathcal{P}_{\ket{g}\bra{g}}\big(\rho(\tilde{t})\big)$,
           for distinguishable systems. For both plots,
           $\Omega\Delta\tilde{t}\approx 0.08$.
           The dots are the results of recursive calculations
           using \eqref{eq:general_rec} and \eqref{prob_written}.
           The solid lines are plots of the damped sinusoid
           \eqref{damped_rabi} that fits the experimental data.
           Recall that $\eta$ is the probability that systems 
           will not suffer a perturbation at the times $n\Delta\tilde{t}$.
           Values of $\eta$ and $\Delta\tilde{t}$ were chosen for
           aesthetics only.
           These plots do look good, but our results are not yet correct.
           See the text.}
  \label{fig:recursive}
\end{figure}

For general $n$, we have
\begin{widetext}
\begin{equation}
\label{eq:general_rec}
p_n(\tilde{t})=\eta\,p_{n-1}(\tilde{t})+
(1-\eta)
\Big(\textrm{cos}^2\big(\Omega\,(\tilde{t}-n\Delta\tilde{t})\big)\, p_{n-1}(n\Delta\tilde{t})
+\textrm{sin}^2\big(\Omega\,(\tilde{t}-n\Delta\tilde{t}) \big)\,
\big(1-p_{n-1}(n\Delta\tilde{t})\big) \Big).
\end{equation}
\end{widetext}
Given \eqref{eq:general_rec}, the predictive probability 
$\mathcal{P}_{\ket{g}\bra{g}}\big(\rho(\tilde{t})\big)$
in \eqref{prob_written} can be calculated recursively.
Figure~\ref{fig:recursive} shows the results of two sample calculations.
The dots are from the recursive calculation using \eqref{eq:general_rec}
with \eqref{prob_written}.
The smooth curves are plots of the decaying sinusoid in \eqref{damped_rabi},
which fits the experimental measurements.
For both figures we have used $\Omega\Delta\tilde{t}\approx 0.08$.
The results in Figure~\ref{rec1} were calculated using $\eta=0.99$
and resulted in a fitted value for the decay factor of
$\gamma/\Omega=0.05$.
The results in Figure~\ref{rec2} were calculated using $\eta=0.997$
and resulted in a fitted value for the decay factor of
$\gamma/\Omega=0.015$.
Recall that $\eta=1$ for a perfectly isolated system.

Our calculation of the predictive probability
results in no frequency shift
away from the Rabi frequency, $\Omega$, at early times,
in agreement with experiments.
Note, however, that our method results in a fitted,
dimensionless decay factor, $\gamma/\Omega$,
that is independent of $\Omega$.
Experiments indicate that the decay factor does in fact
depend on the Rabi frequency~\cite{meekhof_rabi_1996,petta_coherent_2005}.
Though our first attempt has been instructive, it is not correct.
We have included this incorrect result not only because it clearly
illustrates our new method, but also because
the recursive structure of \eqref{eq:general_rec} will
reappear when we work with indistinguishable systems and interference events.
And it will be interesting
to see how the principle of indistinguishability changes the outcome.

We have shown how, simply by distinguishing between parameters 
and coordinates, we are led inevitably to the prediction
of decoherence for systems interacting with an uncontrolled environment.
We have not needed a phase-destroying
master equation or any effective, non-Hermitian Hamiltonian operator.
Even the Born probability \eqref{born_rabi} we have used to produce
Figure~\ref{fig:recursive} is the result of standard calculations and
is found in textbooks.

We have treated our calculation quantum mechanically, but we have also
treated the measured systems and the interference events
as though they are distinguishable.
They are not.
Next we will show that, when we treat systems and interference events
as indistinguishable, our results agree
with experiments, even where calculations using the master equation do not.

\subsection{Predictive Probability and Indistinguishability}
\label{sec:prob_indist}
In real experiments,
useful measurements involve ensemble averages over identically
prepared physical systems.
Different members of the ensemble suffer different but
indistinguishable sequences of environmental interactions.
This is true even for experiments in
which only one physical system is present at a given
time in the lab.
To understand why, consider the result of a hypothetical, active
measurement on any single member of an ensemble of physical systems:
\begin{enumerate}
\item At $\tilde{t}=0$ a physicist actively prepares the systems.
\item At $\tilde{t}=1\Delta\tilde{t}$ subset $A$ is passively prepared
to be in a mixture of the ground state and the excited state.
\item At $\tilde{t}=2\Delta\tilde{t}$ subset $B$ is passively prepared
to be in a mixture of the ground state and the excited state.
\item At some time coordinate value $\tilde{t} > 2\Delta\tilde{t}$
a physicist actively measures a single system from the ensemble
to be in the ground state, $\ket{g}$.
\end{enumerate}
Because of the probabilistic nature of quantum mechanics, it is
in principle impossible to distinguish if the actively
measured system was a member
of subset $A$, subset $B$, both, or neither.
We will address this principle by treating as indistinguishable
the probabilities resulting from different interference events.
And because we cannot distinguish by a measurement any one system
from the others, we will also treat the systems as indistinguishable.

Dealing with indistinguishable systems and probabilities
is an interesting problem.
One must understand what it means to
choose a subset from an ensemble of \emph{indistinguishable} systems, assign
to that subset a preparation time \emph{distinguishable} from the preparation
time of the unperturbed systems, and finally to allow the subset back
into the ensemble where it is again indistinguishable from the rest.

When modeling actual experiments, one would probably use
a full simulation.
To maintain transparency,
however, we will again search for an analytical formula.
Our task is to invent a new, simple transformation
scenario respecting the principle of
indistinguishability outlined above.
Let us create an analogue of \eqref{prob_written} and \eqref{eq:general_rec},
but for indistinguishable systems.

First, assume that we wish to calculate the predictive probability
given that there have been $n$ chances for systems to have
suffered a single interference event.
Assuming again that interference events possibly occur at intervals of
$\Delta\tilde{t}$, the predictive probability can be labeled
$\mathcal{P}_{\ket{g}\bra{g}}\big(\rho(n\Delta\tilde{t})\big)$.
Our scenario is as follows:
\begin{enumerate}
\item\label{choice1} Choose a combination of time intervals
(of length $\Delta\tilde{t}$) such that systems have
survived only a duration of $1\,\Delta\tilde{t}$ before being
passively measured.
\item\label{adj1} Adjust the probability \textbf{at}
$\boldsymbol n \boldsymbol\Delta \tilde{\boldsymbol t}$ accordingly.
\item Because the systems and interference events are indistinguishable,
so are the intervals between the events.
Put the intervals chosen above back into the original set.
\item\label{choice2} Choose a combination of time intervals
such that systems have
survived a duration of $2\,\Delta\tilde{t}$ before being
passively measured.
\item\label{adj2} Adjust the probability \textbf{at}
$\boldsymbol n \boldsymbol\Delta \tilde{\boldsymbol t}$ accordingly.
\item Put the intervals chosen above back into the original set.
\item Repeat until reaching $n\Delta\tilde{t}$.
\end{enumerate}
We have again chosen to parametrize our scenario with a time
scale, $\Delta\tilde{t}$.
However, we will require a slightly different interpretation
for the number parameterizing the systems' susceptibility to environmental interference.
We shall now use $\beta$ (rather than $\eta$), with $0\le \beta\leq 1$ and
$\beta=1$ for a perfectly isolated system.
Physically, $\beta$ will be the probability that a randomly chosen time interval
will have come before an interference event.
This can be better understood after equation \eqref{eq:mom}.

To implement steps \ref{choice1} and \ref{choice2} above,
we will make use of the binomial distribution,
\begin{equation}
\label{eq:bin_dist}
b(n,k,\beta) \equiv
{n \choose k} \beta^k (1-\beta)^{n-k}.
\end{equation}
The distribution in \eqref{eq:bin_dist} gives
the probability for the occurrence of any combination, regardless
of order, of $k$ events with probability $\beta$ and $(n-k)$ events
with probability $(1-\beta)$.
Because the distribution gives probabilities for combinations rather
than permutations, with it we can treat indistinguishable interference
events.
Note also the normalization
\begin{equation}
\label{eq:binorm}
\sum_{k=0}^{n} {n \choose k} \beta^k (1-\beta)^{n-k} = 1.
\end{equation}
Using \eqref{eq:binorm} we will relate the binomial distribution to a probability.

To implement steps \ref{adj1} and \ref{adj2} above, we will once more use the rule
deduced in Section~\ref{sec:ensembles}, to reset to $t=0$
the time evolution parameter
of those systems passively prepared as a result of environmental interference.
To calculate the predictive probability we will therefore
make use of our choice of 
time asymmetric boundary conditions \eqref{eq:hardy_axiom_1}.

It is instructive to write the formula for a specific case.
And for simplicity, we assume that systems in the ensemble will
have suffered at most \emph{only one} interference event before an
active measurement occurs.
This will truncate our formula at a reasonable size, and for
values of $\beta$ close to $1$, we expect this to be a good
approximation at early times.
This is the probability at $4\Delta\tilde{t}$
to find in $\ket{g}$ systems that have been initially
prepared at  $\tilde{t}=0$ to be in $\ket{e}$:
\begin{widetext}
\begin{eqnarray}
\label{eq_simple_case}
\mathcal{P}_{\ket{g}\bra{g}}\big(\rho(4\Delta\tilde{t})\big) & = &
    b(4,4,\beta)\Big(\textrm{cos}^2(\Omega\, 0\Delta\tilde{t})\,\textrm{sin}^2(\Omega\, 4\Delta\tilde{t})
 + \textrm{sin}^2(\Omega\, 0\Delta\tilde{t})\,\textrm{cos}^2(\Omega\, 4\Delta\tilde{t})\Big)+\nonumber \\
& & b(4,3,\beta)\Big(\textrm{cos}^2(\Omega \,1\Delta\tilde{t})\,\textrm{sin}^2(\Omega \,3\Delta\tilde{t})
 + \textrm{sin}^2(\Omega \,1\Delta\tilde{t})\,\textrm{cos}^2(\Omega \,3\Delta\tilde{t})\Big)+\nonumber \\
& & b(4,2,\beta)\Big(\textrm{cos}^2(\Omega \,2\Delta\tilde{t})\,\textrm{sin}^2(\Omega \,2\Delta\tilde{t})
 + \textrm{sin}^2(\Omega \,2\Delta\tilde{t})\,\textrm{cos}^2(\Omega \,2\Delta\tilde{t})\Big)+\nonumber \\
& & b(4,1,\beta)\Big(\textrm{cos}^2(\Omega \,3\Delta\tilde{t})\,\textrm{sin}^2(\Omega \,1\Delta\tilde{t})
 + \textrm{sin}^2(\Omega \,3\Delta\tilde{t})\,\textrm{cos}^2(\Omega \,1\Delta\tilde{t})\Big)+\nonumber \\
& & b(4,0,\beta)\Big(\textrm{cos}^2(\Omega \,4\Delta\tilde{t})\,\textrm{sin}^2(\Omega \,0\Delta\tilde{t})
 + \textrm{sin}^2(\Omega \,4\Delta\tilde{t})\,\textrm{cos}^2(\Omega \,0\Delta\tilde{t})\Big).
\end{eqnarray}
\end{widetext}

The explanation of \eqref{eq_simple_case} is straightforward.
We need to relate the binomial distribution to the passage of time, so
at every step $k$, with $0\leq k\leq n=4$, we will count the 
(normalized) number of combinations for arranging the $n$
time intervals such that $k$ of them came before an interference event.
This number is given by $b(n,k,\beta)$.
Because possible interference events occur at increments of the time
scale, $\Delta\tilde{t}$, the weight $b(n,k,\beta)$ must then be attached
to any passive preparation occurring at $k\Delta\tilde{t}$.

To understand the effects of a passive preparation occurring
at $k\Delta\tilde{t}$,
look at the second line of \eqref{eq_simple_case}, which corresponds to $k=3$.
For the terms in this line,
the only possible interference event occurred at $3\Delta\tilde{t}$.
The probability for a system, prepared initially by the physicist
to be in the excited state, $\ket{e}$, to be passively prepared to
be in the ground state, $\ket{g}$, is
$\textrm{sin}^2(\Omega \,3\Delta\tilde{t})$.
According to our rule for resetting the time parameter,
the probability at $4\Delta\tilde{t}$ to find that system again in
the ground state is 
$\textrm{cos}^2(\Omega \,1\Delta\tilde{t})$.
The argument of the cosine term is $\Omega \,1\Delta\tilde{t}$
because, relative to $4\Delta\tilde{t}$,
the system will have had its time parameter reset to $t=0$ a
duration of $(4-3)\Delta\tilde{t}=1\Delta\tilde{t}$ earlier.
Therefore,
\begin{equation*}
\textrm{cos}^2(\Omega \,1\Delta\tilde{t})\,\textrm{sin}^2(\Omega \,3\Delta\tilde{t})
\end{equation*}
is the probability that a system at $4\Delta\tilde{t}$ will be measured in
$\ket{g}$ given that it was passively prepared at $3\Delta\tilde{t}$
to be in $\ket{g}$ and actively prepared by the experimenter at $\tilde{t}=0$
to be in $\ket{e}$.
Likewise, the other term in the second line of \eqref{eq_simple_case},
\begin{equation*}
\textrm{sin}^2(\Omega \,1\Delta\tilde{t})\,\textrm{cos}^2(\Omega \,3\Delta\tilde{t})
\end{equation*}
is the probability that a system at $4\Delta\tilde{t}$ will be measured in
$\ket{g}$ given that it was passively prepared at $3\Delta\tilde{t}$
to be in $\ket{e}$ and actively prepared by the experimenter at $\tilde{t}=0$
to be in $\ket{e}$.

Let us introduce the notation
$\mathcal{P}^{(i)}_{\ket{g}\bra{g}}\big(\rho(n\Delta\tilde{t})\big)$
to represent the predictive probability under the assumption
that systems on average will have suffered at most
$i$ interference events before measurement.
Then for general $n$,
\begin{widetext}
\begin{equation}
\label{eq:one_event}
\mathcal{P}^{(1)}_{\ket{g}\bra{g}}\big(\rho(n\Delta\tilde{t})\big) = \sum_{k=0}^n
b(n,k,\beta)\Big(\textrm{cos}^2(\Omega(n-k)\Delta\tilde{t})\,\textrm{sin}^2(\Omega\,k \Delta\tilde{t})
 + \textrm{sin}^2(\Omega(n-k)\Delta\tilde{t})\,\textrm{cos}^2(\Omega\,k\Delta\tilde{t})\Big).
\end{equation}
\end{widetext}
By simply exchanging the
$\textrm{cos}^2(\Omega\, k\Delta\tilde{t})$ and
$\textrm{sin}^2(\Omega\, k\Delta\tilde{t})$ terms,
we calculate
$\mathcal{P}^{(1)}_{\ket{e}\bra{e}}\big(\rho(n\Delta\tilde{t})\big)$,
which is the probability to find the systems
in the excited state, $\ket{e}$.

In \eqref{eq:one_event},
we have assumed that systems will have suffered at most one interference event.
To allow for the possibility of multiple events, the terms
$\textrm{sin}^2(\Omega\, k\Delta\tilde{t})$ and
$\textrm{cos}^2(\Omega\, k\Delta\tilde{t})$ 
must be replaced with new functions of $k\Delta\tilde{t}$,
that predict the effects of interference events prior to the
single event assumed in \eqref{eq:one_event}.
For general $i$,
\begin{widetext}
\begin{equation}
\label{eq:two_events}
\mathcal{P}^{(i)}_{\ket{g}\bra{g}}\big(\rho(n\Delta\tilde{t})\big) = \sum_{k=0}^n
b(n,k,\beta)\Big(\textrm{cos}^2(\Omega(n-k)\Delta\tilde{t})\,
\mathcal{P}^{(i-1)}_{\ket{g}\bra{g}}\big(\rho(k\Delta\tilde{t})\big)
 + \textrm{sin}^2(\Omega(n-k)\Delta\tilde{t})\,
\mathcal{P}^{(i-1)}_{\ket{e}\bra{e}}\big(\rho(k\Delta\tilde{t})\big)
\Big).
\end{equation}
\end{widetext}

In \eqref{eq:two_events}, the recursive structure of
\eqref{prob_written} and \eqref{eq:general_rec} has reappeared,
but here our systems and interference events are indistinguishable.
Handling the possibility for more and more interference events requires
the nesting of more and more summed terms into \eqref{eq:two_events}.
These nested equations are significantly more difficult to solve
than are the recursive equations used for distinguishable systems.

We have written the predictive probability as a function of $n\Delta\tilde{t}$.
The final step is to scale our result back to the time coordinate, $\tilde{t}$.
The first moment of the binomial distribution is
\begin{equation}
\label{eq:mom}
\langle k \rangle = \sum_{k=0}^{n} {n \choose k} \beta^k (1-\beta)^{n-k} \,k =\beta\, n.
\end{equation}
After stepping through time to $n\Delta\tilde{t}$,
on average $\beta\,n$ of the intervals will have preceded
the interference event number $i$.
This provides us a physical interpretation of our two parameters,
and to ensure that the time scales with something physical, we will need to use
$\langle k \rangle \Delta\tilde{t}= \beta\, n\Delta\tilde{t}=\tilde{t}$.
After a calculation of the predictive probability as a function of
$n\Delta\tilde{t}$, we make the replacement
\begin{equation}
n\rightarrow \frac{\tilde{t}}{\beta \,\Delta\tilde{t}}\,.
\end{equation}
This restricts us to non-zero values of $\beta$ and $\Delta\tilde{t}$.
We have also simply interpolated between the discrete
values of $n$ at which \eqref{eq:two_events} is actually defined.
Because our time scale is understood to be an average value,
it would be inappropriate to assume anything more complicated.

\begin{figure}[h]
\includegraphics[width=.5\textwidth]{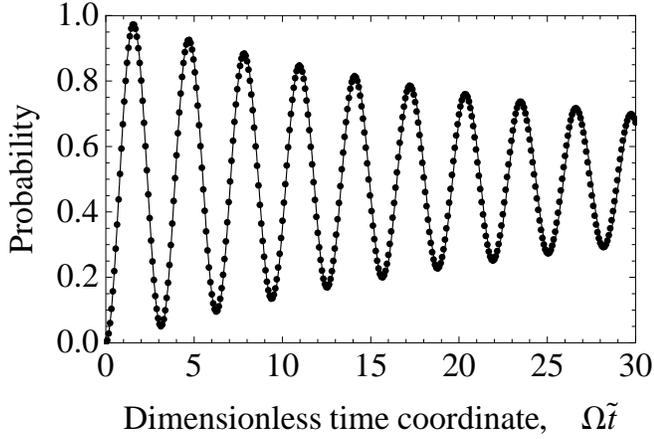}
\caption{Plot of the predictive probability,
$\mathcal{P}^{(5)}_{\ket{g}\bra{g}}\big(\rho(\tilde{t})\big)$,
for indistinguishable systems.
For the dots we have used equations \eqref{eq:two_events}
and \eqref{eq:subst}.
The solid line is a plot of the damped sinusoid \eqref{damped_rabi}
that fits the experimental data.
We have used $\Omega\Delta\tilde{t}\approx 0.7$ and $\beta=0.995$,
and we have fit $\gamma / \Omega = 0.039$.}
\label{fig:indis}
\end{figure}

The predictive probability at $\tilde{t}$,
assuming $i$ interference events, to find 
in the ground state systems
initially prepared in the excited state is therefore
\begin{equation}
\label{eq:subst}
\mathcal{P}^{(i)}_{\ket{g}\bra{g}}\big(\rho(\tilde{t})\big)=
\mathcal{P}^{(i)}_{\ket{g}\bra{g}}\big(\rho(n\Delta\tilde{t})\big),\quad
n\rightarrow\frac{\tilde{t}}{\beta\,\Delta\tilde{t}}\,.
\end{equation}
Numerical solution of \eqref{eq:two_events} is straightforward.
One can also perform the summations
and find closed form expressions for the predictive probabilities.
In Figure \ref{fig:indis} we have plotted 
$\mathcal{P}^{(5)}_{\ket{g}\bra{g}}\big(\rho(\tilde{t})\big)$,
which is the predictive probability given that there have been 
at most $5$ interference events.
For clean experiments, with $\beta$ close to $1$,
we expect our result to be a good approximation,
especially at short times.
Again, the agreement with the experimentally measured damped sinusoid is
quite good.

As opposed to the previous results for distinguishable systems
(see Section \ref{sec:dist_dev}), however,
when we treat the systems and interference events as indistinguishable, we
find that the damping factor, $\gamma$, does in general depend on the 
Rabi frequency, $\Omega$.
In the next section we shall compare 
$\mathcal{P}^{(5)}_{\ket{g}\bra{g}}\big(\rho(\tilde{t})\big)$ with measurements.
But if we allow a rather crude approximation, we can find the 
form of the dependence
of $\gamma$ on $\Omega$ for $\beta$ very close to 1 and small $\Delta\tilde{t}$.
For $\beta\approx 1$, the predictive probability is dominated by the
terms proportional to $b(n,k=n,\beta)$.
Solving the truncated form
\begin{equation}
\mathcal{P}_{\ket{g}\bra{g}}\big(\rho(n\Delta\tilde{t})\big) \approx \sum_{k=0}^n
b(n,k,\beta)\textrm{sin}^2(\Omega\,k\Delta\tilde{t})
\end{equation}
and using \eqref{eq:subst}, we get
\begin{eqnarray}
\mathcal{P}_{\ket{g}\bra{g}}\big(\rho(\tilde{t})\big)& \approx & 
\frac{1}{4} \Big(2-
\big( 1-\beta (1-e^{-2 i \Delta\tilde{t}\Omega})\big)^\frac{\tilde{t}}{\beta\Delta\tilde{t}}
\nonumber \\
\label{eq:approx}
& &\,\,\quad-\big( 1-\beta
(1-e^{+2 i \Delta\tilde{t}\Omega})\big)^\frac{\tilde{t}}{\beta\Delta\tilde{t}}
\Big).
\end{eqnarray}
For small $\Delta\tilde{t}$, \eqref{eq:approx} is
\begin{equation}
\mathcal{P}_{\ket{g}\bra{g}}\big(\rho(\tilde{t})\big)=
\frac{1}{2}\Big(1-e^{-\gamma \tilde{t}}\,\big(\textrm{cos}(2 \Omega \tilde{t})
+\textrm{O}(\Delta\tilde{t}^{\,2})\big)\Big),
\end{equation}
where
\begin{equation}
\gamma = 2\, (1-\beta)\, \Omega^2\, \Delta\tilde{t} + \textrm{O}(\Delta\tilde{t}^{\,3}).
\end{equation}
For very clean systems and with small time scales for interference,
$\gamma$ is quadratic in $\Omega$.
Though we explicitly see dependence on the Rabi frequency, this
is only an approximation.
We will not use it to compare with experimental results.

\subsection{Comparison With Experiment and With the Standard Theory}
\label{sec:compare}
The standard method for calculating a probability for
decoherence in dynamically evolving systems
is to use the quantum master equation.
It is the method of choice when one works
without time asymmetric boundary conditions and does not distinguish
between parameters, $t$, and coordinates, $\tilde{t}$.
The generic solution of the master equation describing a system
undergoing Rabi oscillations and \emph{on resonance}
is~\cite{petruccione_open_quantum_systems}
\begin{equation}
\label{eq:mast_sol}
\mathcal{P}^{ME}_{\ket{g}\bra{g}}(t)=\frac{4\Omega^2}{\Gamma^2+8\Omega^2}
\big(1-e^{-3 \Gamma t/4}(\textrm{cos}\, \mu t + \frac{3\Gamma}{4 \mu}
\textrm{sin}\,\mu t)  \big).
\end{equation}
To match our convention, we have added a factor of $2$ to the
definition of Rabi Frequency in~\cite{petruccione_open_quantum_systems}.
In \eqref{eq:mast_sol}, the $t$ is a parameter rather than a coordinate, and
$\mu=\sqrt{4\Omega^2-\big( \frac{\Gamma}{4} \big)^2}$.
Unless a specific environmental interaction has been assumed,
$\Gamma$ is the spontaneous emission rate
for the system in its excited state.
It is in general a constant, independent of the Rabi frequency.

A significant shift from the Rabi frequency, $\Omega$, is not
observed in experiments.
If one assumes very strong driving, $2\Omega\gg \Gamma/4$, the limit
of \eqref{eq:mast_sol} is
\begin{equation}
\label{eq:mast_lim}
\mathcal{P}^{ME}_{\ket{g}\bra{g}}(t)= \frac{1}{2} \big(
1-e^{-3 \Gamma t/4} \textrm{cos}\,2\Omega t \big).
\end{equation}

The solution \eqref{eq:mast_lim} of the quantum master equation,
however, is still not correct.
Experiments reveal that the damping factor, $\gamma$, does depend
on the Rabi frequency, $\Omega$.
In one experiment~\cite{meekhof_rabi_1996},
Rabi oscillations are observed between a series of different
levels of a single $^9\textrm{Be}^+$ ion.
The different levels are described by the kets
$\ket{\downarrow, n}$ and $\ket{\uparrow, n+1}$, where
$\ket{\downarrow}$ and $\ket{\uparrow}$ are internal states of the Be ion, and
$\ket{n}$ represents vibrational Fock states.
Rabi oscillations are measured for the
frequencies~\cite{wineland_experimental_1998,meekhof_rabi_1996}
\begin{equation}
\label{eq:freq_seq}
\Omega_{n,n+1} = \Omega \frac{0.202 \,e^{-0.202^2/2}}{\sqrt{n+1}} L^1_n(0.202^2),
\end{equation}
where $L^1_n$ is the generalized Laguerre polynomial.
The corresponding damping factor, $\gamma_n$, is measured~\cite{meekhof_rabi_1996}
to increase with $n$ according to 
\begin{equation}
\label{eqn:gamma_meas}
\frac{\gamma_n}{\gamma_0}\approx (1+n)^{0.7}. \qquad \textrm{(Measured)}
\end{equation}

The measured result \eqref{eqn:gamma_meas} has been very puzzling.
Without assuming any specific character for environmental interference,
the solution \eqref{eq:mast_lim} of the master equation predicts a constant
\begin{equation}
\label{eqn:gamma_meq}
\frac{\gamma_n}{\gamma_0}=1. \qquad\quad \textrm{(Master equation)}
\end{equation}
Of course, specific choices for the interaction operator
for the master equation can lead to damping factors that depend on
the Rabi frequency.
In~\cite{schneider_decoherence_1998}, the effects of intensity
fluctuations of the driving laser are studied.
Solving the master equation results in a damping factor that does depend
on the Rabi frequency, but the calculated exponent in the prediction for
\eqref{eqn:gamma_meas} is $\frac{1}{2}$ rather than the measured $0.7$.
In~\cite{murao_decoherence_1998}, 
the effects of imperfect dipole transitions and of fluctuations of the trap potential
are both studied, but solution of the master equation does not result in
the correct relation for damping factors unless, at the end of the calculation,
$\gamma_n$ is actually postulated to depend on
$(\Omega_n)^d$, where $d$ is an exponent not limited to the set of integers.
To match \eqref{eqn:gamma_meas}, $d$ must be tuned to
$0.4$ or  $2.4$, depending
on the character of the environmental coupling.

Studies using
the standard formalism of the master equation suggest that the observed
dependence of the damping factor on the Rabi frequency
is not common and should only be expected in those experiments
on systems suffering perturbations of precisely the correct character.
And even then, the prediction is either not correct or else a correct result is
postulated at the end of the calculation.

In another experiment~\cite{petta_coherent_2005}
with spin states of two electrons in a double quantum dot,
the decay factor, $\gamma$, is stated to be proportional to the
Rabi frequency, $\Omega$, though no mathematical relation is given.
Such a result for a system so fundamentally different from the previous
one suggests that some dependence of $\gamma$ on $\Omega$ is general,
in agreement with our approach.

It should be noted that,
in the theoretical investigation~\cite{romito_decoherence_2007}
of this quantum dot experiment, the authors have used the master equation to find
that $\gamma$ can depend on quantum state fluctuations
resulting from the tunneling of electrons between quantum dots.
The presence of tunneling
is very reasonable~\cite{hu_charge_fluctuation_induced_2006}, and 
it also changes the value of $\Omega$.
The damping factor and the Rabi frequency can be thereby related.
The study~\cite{romito_decoherence_2007} also reveals that the effect
of coupling to a bosonic reservoir alone is not sufficient to predict
a dependence of $\gamma$ on $\Omega$.

\begin{figure}[h]
\includegraphics[width=.5\textwidth]{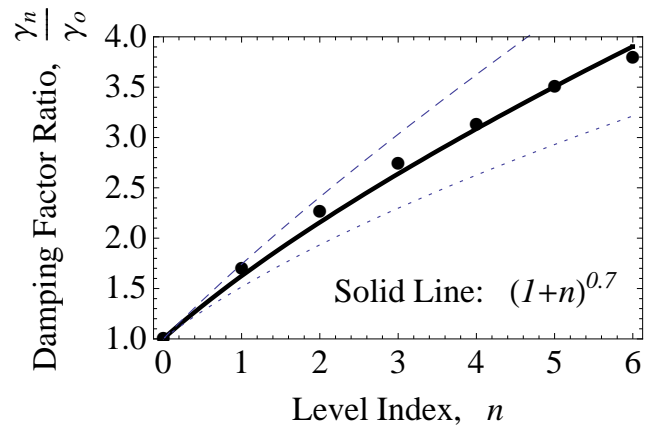}
\caption{Matching the experimental results for the ratio of
damping factors, $\frac{\gamma_n}{\gamma_0}$.
The large dots result from our theoretical calculation of the predictive
probability using \eqref{eq:two_events} and \eqref{eq:subst}.
The solid line is the experimentally measured relation, $(1+n)^{0.7}$.
To indicate a scale for the exponent,
the thin dashed line is a plot of $(1+n)^{0.8}$,
and the thin dotted line is a plot of $(1+n)^{0.6}$.}
\label{fig:fit}
\end{figure}

As mentioned above,
by assuming only a time scale, $\Delta \tilde{t}$, and a parameter describing systems'
susceptibility to environmental decoherence, $\beta$, our
calculation of a predictive probability \eqref{eq:two_events}
results generally in a damping factor that depends on the Rabi frequency.
And not only is there a dependence, but we
also fit the measured relation in \eqref{eqn:gamma_meas}
without making any further assumptions.
In Figure~\ref{fig:fit} is the result of fitting the decay factor, $\gamma_n$, to 
$\mathcal{P}^{(5)}_{\ket{g}\bra{g}}\big(\rho(\tilde{t})\big)$,
calculated with the sequence of frequencies in \eqref{eq:freq_seq}
and with $\Omega_0\Delta\tilde{t}\approx 0.2$.
(The exponent can be shifted by choosing different time scales.)
As $n$ increases, the fit deteriorates somewhat,
but we do not know over what range the fit was performed.

Figure~\ref{fig:fit} shows a remarkable agreement with experiment.
Not only does our approach predict a general dependence of
$\gamma$ on $\Omega$, but we also get the form correct
without requiring any specific assumptions for the character
of environmental interference.
We conclude that
the experimentally verified dependence of $\gamma$ on
$\Omega$ is a measurable effect
of the indistinguishability of separate, uncontrolled interactions
between quantum systems and their environment.

\section{A New Approach to Quantum Dynamics}
\label{sec:qd}
In the previous sections we have suggested a new way to apply 
quantum mechanics when studying dynamical systems.
Note that we have not broken any of the rules of quantum mechanics.
Based on the difference between $\tilde{t}$ and $t$,
we also draw very naturally the
distinction between what is calculated and what is measured
\eqref{eq_bornshift}.
And when environmental interactions are no longer present, the theory
reduces to standard quantum mechanics, though still with the 
asymmetric boundary conditions \eqref{eq:hardy_axiom_1}
and still with the phenomenologically identified preparation time.

The master equation formalism is developed by finding conditions on
density operators representing physical systems 
and evolving with the parameter
$t$~\cite{petruccione_open_quantum_systems}.
The uncontrolled environment is represented by another density operator,
containing the density operator of the systems,
and also evolving with the same $t$.
(This is, of course, a gross over-simplification of the development.)
When we distinguish between the parametric $t$ and the
time coordinates, $\tilde{t}$, in which measurements are made,
the theoretical derivation of the master equation no longer applies.
In fact, it seems that the master equation formalism may be too restrictive
to predict for systems undergoing Rabi oscillations a general
dependence of damping on frequency.
But there is some similarity between our method and the standard approach.
The use of density matrices and the
theoretical tools for projecting quantum states 
onto a macroscopic environment will be very useful.
We will get nothing for free.
One must always understand the nature of environmental interactions.

Our approach also suggests natural resolutions to some well-known problems.
For instance, the entropy of quantum mechanical systems is 
predicted to be constant in time.
The von Neumann entropy is
\begin{equation}
S\big(\rho(t)\big)\equiv -\Tr\big(\rho(t) \,\textrm{ln}\rho(t) \big).
\end{equation}
The (unitary) time evolution of the density operator is
\begin{equation}
\rho(t)=U^\dagger (t)\rho(0) U(t),
\end{equation}
where $U^\dagger(t)$ is given by \eqref{state_group}.
Thus,
\begin{eqnarray}
\label{eq:const_ent}
S\big(\rho(t)\big)&=& -\Tr\big(U^\dagger (t)\rho(0) U(t)
\,U^\dagger (t)\textrm{ln} \rho(0) U(t) \big) \nonumber \\
&=& S\big(\rho(0)\big).
\end{eqnarray}
Here, the $t$ is a parameter, so \eqref{eq:const_ent} is true
for parametric time evolution.
With our new understanding, physical systems are
represented by $\rho(\tilde{t})$.
When systems are not perfectly isolated,
$\rho(\tilde{t})\neq\rho(t)$.
The appropriate von Neumann entropy is
$S\big(\rho(\tilde{t})\big)$, and \eqref{eq:const_ent} no longer applies.
In the standard formalism, this puzzle is solved with non-unitary time evolution.

When working in the standard formalism and with time symmetric boundary
conditions, one invokes the act of measurement to induce a
fundamentally irreversible ``collapse of the wave function.''
This is equivalent to invoking the passive measurements
of our transformation scenarios.
But with the new approach we can treat these ``collapses''
as sequences of events.
As with classical statistical mechanics, we are free to ask what
is the likelihood that the effects of environmental interference can be reversed
without active intervention by a physicist.
And as with the classical theory, finding an answer requires
nothing but a counting exercise.
This is very satisfying intuitively.

The time translational invariance of quantum systems is also addressed in a 
natural way.
The value of the time coordinate, $\tilde{t}$, is always physically irrelevant.
Because of our choice of boundary conditions, however, the time parameter value,
$t$, represents a duration.
Durations are physically significant, and, as expected, there is 
in general no invariance for parametric translations.

Perhaps most compelling is the general applicability of our approach.
Here we have concentrated on the quantum mechanics of discrete
spectra and Rabi oscillations experiments.
In a forthcoming paper we find the choice of time asymmetric
boundary conditions \eqref{eq:hardy_axiom_1} also to be quite
powerful when used to derive the equations of non-relativistic scattering theory.

\section{Conclusion}
We have investigated a new approach to quantum dynamics.
It is based on the phenomenologically motivated choice of
time asymmetric boundary conditions for the Schr\"odinger equation.
The correct application of the intrinsically
time asymmetric theory requires one to distinguish between time
evolution parameters, $t$, and time coordinates, $\tilde{t}$.
Remarkably, for systems suffering environmental interference, predictions
of the experimental signatures of an \emph{extrinsic} arrow of time
are unavoidable consequences of the \emph{intrinsic} time asymmetry.

Clearly this theory needs to be developed formally.
It is different from the standard formalism of
the phase-destroying master equation and time evolution driven by
effective, non-Hermitian Hamiltonian operators.
When we distinguish between time coordinates and time parameters,
the standard formalism, which is derived for parametric evolution only,
does not apply.

As an application, we have matched a dynamical
measurement performed on systems undergoing Rabi oscillations.
Investigations using the master equation have not produced correct results.
Furthermore, we conclude that the measured dependence of a damping
factor on the Rabi frequency is a consequence of the indistinguishability
of separate interactions between quantum systems and their environment.

\end{document}